\documentclass[12pt]{iopart}
\usepackage{iopams}  
\usepackage{amsthm}

\def\thorn{\hbox{\rm I}\kern-0.32em\raise0.35ex\hbox{\it o}}
\def\edth{\hbox{$\partial$\kern-0.25em\raise0.6ex\hbox{\rm\char'40}}}
\def\thornprime{\thorn^\prime}
\def\edthprime{\edth^\prime}
\newtheorem{lemma}{Lemma}

\begin{document}

\title[\bf Petrov D vacuum spaces revisited: Identities and Invariant Classification]{\bf Petrov D vacuum spaces revisited: Identities and Invariant Classification}

\author{S. Brian Edgar$^1$, Alfonso Garc\'{\i}a-Parrado G\'omez-Lobo$^1$\footnote{Present address: Ghent University, Department of Mathematical Analysis. Galglaan 2, 9000 Ghent, Belgium.} and 
Jos\'e M. Mart\'{\i}n-Garc\'{\i}a$^{1,2,3}$}

\address{$^1$ Matematiska institutionen, Link\"opings Universitet,\\
SE-581 83 Link\"oping, Sweden.}
\address{$^2$ Laboratoire Univers et Th\'eories, 
Observatoire de Paris, CNRS, \\
Univ. Paris Diderot, 5 place Jules Janssen, 92190 Meudon, France.
}
\address{$^3$
Institut d'Astrophysique de Paris, Univ. Pierre et Marie Curie, \\
CNRS, 98$^{bis}$ boulevard Arago, 75014 Paris, France.}
\eads{\mailto{bredg@mai.liu.se}, \mailto{algar@mai.liu.se},
\mailto{Jose.Martin-Garcia@obspm.fr}}

\begin{abstract}
For Petrov D vacuum spaces, two simple identities are rederived and some new identities 
are obtained, in a manageable form, by a systematic and transparent analysis using the GHP 
formalism. This gives a complete involutive set of tables  for the four GHP derivatives on each of  the 
four GHP spin coefficients and the one Weyl tensor component.  It follows directly from 
these results that  the 
theoretical upper bound on the order of covariant differentiation of the Riemann tensor required for a Karlhede classification of these spaces  is  reduced to two.
\end{abstract}

\pacs{02.40.-k, 04.20.-q, 04.20.Jb}


\section{Introduction}

In his  pioneering  work in the NP formalism,  which provided definitive results on vacuum Type D 
spacetimes \cite{kin}, surprisingly Kinnersley did not exploit, and in fact seems to have been unaware of,  two simple NP identities for these spacetimes
\begin{eqnarray}
I_1\equiv \pi\bar\pi -\tau\bar\tau  =0= \rho\bar \mu - \bar \rho\mu   \equiv I_2
\label{II1I2}
\end{eqnarray}
(In the usual NP notation \cite{np}, the two tetrad vectors $l,n$ are principal null vectors of the Weyl tensor.)
More surprisingly, it seems that it took longer than another decade   before these identities were noticed; and, even more surprisingly, they were obtained by chance --- for the vacuum case. Debever and McLenaghan  \cite{demc} had discovered, as the result of a simple calculation, for Type D {\it electrovac} solutions  that\begin{eqnarray}
\Phi( \pi\bar\pi -\tau\bar\tau )  =0= \Phi(\rho\bar \mu - \bar \rho\mu)
\label{II1I2e}
\end{eqnarray}
from which (\ref{II1I2}) immediately follows --- {\it but only for $\Phi \ne 0$}; 
this led them to investigate if these identities were also true for the vacuum case, and they  confirmed the validity of  identities (\ref{II1I2})
 for the {\it vacuum} case --- by a direct calculation {\it using the explicit metrics} given by Kinnersley. 
 Subsequently, Czapor and McLenaghan \cite{czmc2} established (\ref{II1I2}) for  Type D 
{\it vacuum} spacetimes --- without integrating the equations, but by a much more complicated 
and much longer calculation than for the electrovac case;   the calculation was only achieved with 
computer support.  Later these authors used this calculation as an example of their Maple Package for 
calculations in the NP formalism \cite{czmc1}.  They pointed out that the presentation of some of the 
intermediate expressions used in their calculations required pages, and that some of the steps 'would be 
virtually impossible by hand'; furthermore, they stressed that 'even aided by computer, this calculation is 
far from automatic'. Their calculation was built on the unusual tetrad gauge choice $\pi = \rho$, using a programme involving  repeated applications of the commutator equations, judicial substitutions and tricky factorisations.

However, the identities (\ref{II1I2}) are clearly  invariant under spin and boost gauge choices, 
and so are valid in any gauge; the GHP formalism \cite{ghp},  \cite{held1}, \cite{edGHP0} which is invariant under such gauge choices, should be a more efficient tool, and it is to be expected that the results should come out without the need for such long calculations, and indeed more transparently than in the NP formalism.  The GHP formalism has proved very efficient
 in a number of calculations involving Type D spaces, \cite{held2}, 
\cite{sw}, \cite{edGHP}, \cite{edlud3},\cite{car1}; indeed, 
in the latter,  Carminati and Vu have re-established these identities   (\ref{II1I2}) as an example of their Maple Package 
for calculations in the GHP formalism. 

These identities do not seem to be well known; for example, an investigation  of the invariant classification
 of vacuum Type D metrics in \cite{cdV1,cdV2} did not use them in a situation where they would have been 
particularly useful. So the first purpose of this paper is to highlight these identities (\ref{II1I2}), 
as well as to draw attention to some other identities  for spin coefficients, some of which we believe to be new.  The second purpose is to emphasise the efficiency of the 
GHP formalism by obtaining  all of these identities   in a comparatively brief, straightforward,  and 
systematic manner within the GHP formalism --- by hand for transparency, but with computer support for accuracy.
 The third purpose is to reconcile the theoretical  and measured upper bounds in the invariant classification for
 Type D vacuum spaces  --- without explicitly integrating the equations.

A  procedure,  for the invariant classification  of spacetimes, which originated with Cartan \cite{car}, was developed by Brans \cite{bra}, and 
refined by Karlhede \cite{kar}, has established theoretical upper bounds on the order of the Cartan scalars
 --- the Riemann tensor and its covariant derivatives, calculated in a particular frame. At first sight, for some Petrov types, these 
upper bounds can be  as high as {\it seven}, so it is of great practical significance to find whether these values are sharp, or whether they  can be lowered to a more manageable number. In 
\cite{cdV1,cdV2} it was shown, for Type D vacuum spacetimes, how the GHP operators could replace the covariant 
derivatives  in the investigation of the different orders of derivatives of the Weyl tensor. By this means, the 
GHP formalism was exploited to lower   the theoretical upper bound to three in the invariant classification of Type D 
spacetimes --- without explicitly integrating the equations; more precisely, it was shown in  \cite{cdV1,cdV2} that the theoretical upper bound was {\it two} for two subcases, while for a third subcase
the bound was shown to be three, but   
 a fourth 
 subcase of Petrov D vacuum spacetimes was
overlooked\footnote{The investigation in \cite{cdV1,cdV2} relied on a result in \cite{fla}
which claimed that there were only two additional subcases in addition to the generic subcase $\rho\rho'\tau\tau' \ne 0 $; this ignored the null orbit possibility $\rho \ne 0, \rho' = 0$ \cite{czmc2}, which yields a subset of the Kinnersley Case IIE metrics, which we call Class II in our Appendix.}.  Petrov type D vacuum is the only Petrov subclass where all spacetimes are known explicitly, 
and the upper bound has been calculated {\it directly from these metrics}
to be {\it two} \cite{ak}, \cite{aa}. In this paper we shall strengthen the result in \cite{cdV1,cdV2}, and reduce  the   upper
bound  to {\it  two} directly for the remaining subcases,  by a further exploitation of the GHP formalism, but without explicitly
integrating the equations; so the theoretical and measured upper bounds are now in
agreement.  To achieve this, we exploit a series of  identities for vacuum Petrov D
spaces, culminating in a complete set of  tables for  each spin coefficient and the one Weyl tensor component, $\Psi_2$. These
complete tables equate the four GHP derivative operators of each of the spin
coefficients and $\Psi_2$  to  expressions built only from the spin
coefficients and $\Psi_2$. 

The paper is organised as follows: the complete tables for the spin coefficients and $\Psi_2$ are computed in section \ref{tables}. In section \ref{classification} we build on the work of \cite{cdV1,cdV2} and show how these results enable us to derive that the Karlhede bound for the vacuum Petrov type D spacetimes  is two, thus improving the result in \cite{cdV1,cdV2}. Our conclusions and a brief discussion are given in the final section. 
The appendix contains a classification of vacuum Petrov type D spacetimes adapted to our purposes.    

The computer calculations alluded to above have been performed with {\em xAct} \cite{xact}. This is a suite of {\em Mathematica} packages with many features, including the canonicalisation of tensor expressions by means of powerful algorithms based on permutation group theory \cite{xperm}, and the ability of working simultaneously with several different frames.

\section{Identities and complete tables}
\label{tables}
 We follow the conventions and notation of \cite{ghp} in our presentation.   
 We will first write down  the basic equations --- the Ricci, Bianchi and commutator equations --- of the GHP formalism particularised, in the usual way, for vacuum, type D spacetimes.    
Because of the simple structure of these spacetimes in the GHP formalism, all equations have $ '$, $^* $ and $^{*}{}' $ counterparts (some of which may not be independent) but we shall not use $^ *$ and $^{*}{}' $ explicitly in the presentation, although they are very useful for checking.  

\medskip
\noindent
{\em Nonvanishing weighted spin coefficients and curvature components.}
$$
\mathcal{S}\equiv\{\rho,\quad\tau,\quad\rho',\quad\tau',\quad\Psi_2\}.
$$

\medskip
\noindent
{\em Ricci equations.}
\begin{eqnarray}
\thorn \rho = \rho^2,\label{ricci1}\\
\edth \rho = \tau(\rho-\bar \rho),\label{ricci2}\\
\thorn\tau = \rho(\tau - \bar \tau '),\label{ricci3}\\
\edth \tau = \tau^2,\label{ricci4}\\
\thornprime \rho = \rho \bar \rho ' - \tau \bar \tau -\Psi_2 + \edthprime\tau,\label{ricci5}
\end{eqnarray}
and their $'$ counterparts; 

\medskip
\noindent
{\em Bianchi equations.}
\begin{eqnarray}
\thorn \Psi_2 = 3 \rho \Psi_2,\label{bianchi0}\\
\edth \Psi_2 = 3 \tau \Psi_2,
\label{bianchi}
\end{eqnarray}
and their $'$  counterparts;

\medskip
\noindent
{\em Commutator equations.} 
\begin{eqnarray}[\thorn \thornprime - \thornprime\thorn] \eta_{pq}   = \Bigl((\bar \tau-\tau')\edth +  ( \tau-\bar \tau')\edth' -p(\Psi_2-\tau\tau') -q(\bar \Psi_2-\bar\tau \bar\tau') \Bigr ) \eta_{pq},  \label{comm1} 
\\ 
\ [\thorn \edth - \edth\thorn] \eta_{pq}   =   \Bigl(\bar \rho\edth -  \bar \tau'\thorn  +q \bar \rho \bar\tau'\Bigr ) \eta_{pq},  \label{comm2} 
\\ 
\ [\thorn \edth' - \edth'\thorn] \eta_{pq}   =   \Bigl(\rho\edth' - \tau'\thorn +p \rho\tau' \Bigr ) \eta_{pq},  \label{comm3}
\\ 
\ [\thorn' \edth - \edth\thorn'] \eta_{pq}   =   \Bigl( \rho'\edth -   \tau\thorn'  - p \rho' \tau \Bigr ) \eta_{pq},  \label{comm4} 
\\ 
\ [\thorn' \edth' - \edth'\thorn'] \eta_{pq}   =   \Bigl(\bar \rho'\edth' -  \bar \tau\thorn' - q \bar \rho' \bar \tau \Bigr ) \eta_{pq},  \label{comm5} 
\\ \ [\edth \edthprime - \edthprime\edth]\eta_{pq} =  \Bigl((\bar \rho'-\rho')\thorn +  (\rho-\bar\rho)\thorn' +p(\Psi_2+\rho\rho') -q(\bar\Psi_2+\bar\rho\bar\rho') \Bigr ) \eta_{pq},  \label{comm6} 
\end{eqnarray}
where we have written out all commutators explicitly acting on a scalar $\eta_{pq}$ of spin and boost weight $(p-q)/2, \ (p+q)/2$ respectively.

\medskip
When we substitute the  Bianchi equations into the commutators for $\Psi_2$ and also make use of the Ricci equations, in some cases we obtain the trivial identity; from the other   commutators we  obtain   the  

\medskip
\noindent
{\em Post-Bianchi equations.}
\begin{eqnarray}\Psi_2(-\thorn'\rho+\thorn \rho' - \tau \bar \tau +\tau' \bar \tau')=0,\label{Ppostb1} \nonumber\\
\Psi_2(\edthprime \rho - \thorn \tau')=0,\nonumber
\label{Ppostb2}
\end{eqnarray}
and their $'$ counterparts; and since $\Psi_2\ne 0$, it follows that
\begin{eqnarray}\thorn'\rho=\thorn \rho' - \tau \bar \tau +\tau' \bar \tau',\label{postb1} \\
\edthprime \rho = \thorn \tau',
\label{postb2}
\end{eqnarray}
and their $'$ counterparts. Both these equations are quoted in \cite{car1}, but the first (\ref{postb1}) is overlooked in \cite{cdV1,cdV2}. 

Applying the commutators to the spin coefficients given in (\ref{ricci1}), (\ref{ricci2}),  (\ref{ricci3}), (\ref{ricci4}), yields no new information.

Note that although we now have equations with terms  involving all four operators acting on $\rho, \tau$ (and hence also, by $'$ symmetry, on $\rho ', \tau '$) these are not all given in separate equations. Therefore  it will be convenient to introduce new symbols; we follow the notation of Carminati and Vu \cite{car1} and  introduce the zero-weighted $Z_1$ and the $\{0,2\}$-weighted $Z_2$,
\begin{eqnarray}
  Z_1 & = & \edthprime \tau,
  \nonumber\\ & = & \thornprime \rho - \rho \bar \rho ' + 
\tau \bar \tau +\Psi_2,\nonumber\\
 & = & \thorn \rho' - \rho \bar \rho' + \tau' \bar \tau'  +\Psi_2, \nonumber\\
  &  =  & \edth \tau' +\bar\rho\rho' - \rho \bar\rho', \label{Zdefn1} 
  \end{eqnarray}
  and
  \begin{eqnarray}
 Z_2 & = &\edthprime \rho, \nonumber \\ & = &\thorn \tau',\label{Zdefn2}
\end{eqnarray}
where we have used the Ricci equations and  post-Bianchi equations to obtain the different alternatives.

Note that, because of  the Ricci and post-Bianchi equations,  
\begin{eqnarray}
Z_1' = Z_1 +\bar \rho'\rho-\rho'\bar\rho, \label{Z1ppty}
\end{eqnarray}
whereas
\begin{eqnarray}
 Z_2'  & = & \edth \rho', \nonumber\\ &=&  \thorn'\tau.\label{Zdefn3}
\end{eqnarray}
We will use $-Z_2'$ in place of the $Z_3$ used in \cite{car1}. Using the Sachs symmetry ${}^*$ we have the properties
\begin{eqnarray}
{{Z_2}^*} ' = Z_2,\label{Z2*}\\
{{Z_1}^*}  = Z_1 + \rho \bar \rho ' - \tau \bar \tau -\Psi_2 .\label{Z1*} 
\end{eqnarray}

We now have a set of equations involving all four operators acting on $\rho, \tau$ and  by $'$ symmetry, on $\rho'$, $\tau'$. These are the Ricci equations, the post-Bianchi equations and (\ref{Zdefn1})-(\ref{Zdefn2}). Thus we can now apply each of the commutator equations to each of the four spin coefficients. 
In a number of cases the commutators are trivially satisfied via the 
Ricci equations, Bianchi equations and post-Bianchi equations. The remaining non-trivial results from the commutators  are  a set of first-order equations for the new symbols
$Z_1$, $Z_2$:
\begin{eqnarray}
\thorn Z_1 =  \rho (2 Z_1 -\bar Z_1 + \tau' \bar\tau' -\bar\rho\rho' +\bar \rho' \rho)  +(\tau - \bar\tau')Z_2,\label{Zfode}\\
\edth Z_1 = \tau (2Z_1 +\Psi_2 + \bar\Psi_2 + \rho\bar\rho'  + \bar\rho'\bar\rho) -(\bar\rho -\rho)Z_2' - \bar\tau'(\rho\bar\rho'  
-\rho \rho' ),  \label{dZ1} 
\end{eqnarray}
and their $'$ counterparts; together with 
\begin{eqnarray}
\thorn Z_2 = 3 \rho Z_2,\nonumber\\
\edth Z_2 = \tau Z_2 +2(\rho - \bar \rho) Z_1 +\bar \rho\Psi_2 -  \rho\bar\Psi _2 + 2\rho\bar\rho'(\rho -\bar\rho),\nonumber\\
\edthprime Z_2 = 3 \tau' Z_2,
\nonumber\\
\thornprime Z_2 = \rho' Z_2 +\bar\Psi_2\tau' +\Psi_2 (\bar \tau -2\tau') -2(\bar\tau- \tau')(Z_1 +\rho\bar\rho'), \label{dZ2} 
\end{eqnarray}
and their $'$ counterparts. 

\smallskip
\qquad              At this stage we have a complete and involutive set of tables for the eight complex quantities in ${\cal S}^{++} = \{\rho, \tau, \rho', \tau', \Psi_2, Z_1, Z_2, Z_2'\}$. 
This enables us to  rederive and complete the result in \cite{cdV1,cdV2}.  All the elements\footnote{As noted above, in \cite{cdV1,cdV2}, the first of the post-Bianchi equations (\ref{postb1}) was overlooked; this is why in that paper the corresponding expressions  for $D^2 \Psi$ and $D^3 \Psi$ contain  {\it nine} variables: there is   an {additional} variable ($\thorn \rho'$), which corresponds to $Z_1'$ in our notation, but which we know is related to $Z_1$ because of (\ref{postb1}).} of ${\cal S}^{++}$ are present at the second order of differentiation of the Weyl tensor, and there are no new quantities introduced at the third order; moreover, there is also no change in the invariance group (including the case overlooked in \cite{cdV1,cdV2}, which corresponds to Class II in the Appendix) at the third order, and  therefore the Karlhede algorithm terminates at third order.

\smallskip





However, in the following, by continuing the systematic analysis, we shall obtain more identities, and then be able to strengthen this result in the next section.

\medskip

The next step is to check the integrability conditions for (\ref{Zfode})-(\ref{dZ1}); i.e., apply each of the commutator equations to   $Z_1$ and substitute the appropriate expressions from (\ref{Zfode})-(\ref{dZ1}). From the  non-trivial results we  obtain   the new equation 
\begin{eqnarray}
\Psi_2(\tau\bar\tau -\tau'\bar\tau' +\rho\bar\rho' - \bar\rho\rho') =0,
\label{ID}
\end{eqnarray}
from which, by taking real and imaginary parts on the second factor, we get the (GHP version of the) identities (\ref{II1I2}):
\begin{eqnarray}
I_1\equiv \tau'\bar\tau' -\tau\bar\tau  =0= \bar\rho\rho' - \rho\bar\rho' \equiv I_2.
\label{I1I2}
\end{eqnarray}
Note that these identities are invariant under the $'$ symmetry (and also the ${}^*$ symmetry). 
These identities cause simplification in some of the post-Bianchi equations (\ref{postb1}),  and in some of the
  definitions (\ref{Zdefn1}) and (\ref{Z1ppty}) (where $Z_1= Z_1'$);  we will make frequent use of these simplifying 
identities in the following.   

In the same way we check the integrability conditions for (\ref{dZ2}); 
 from the  non-trivial results we  obtain   the new equation 
\begin{eqnarray}
\Psi_2 (Z_2 +\rho\bar \tau-2\rho\tau'+\tau'\bar\rho)=0,
\label{PsiZdefn2}
\end{eqnarray}
from which  we obtain $Z_2$,  
\begin{eqnarray}
 Z_2 =- \rho\bar \tau+2\rho\tau'-\tau'\bar\rho
\label{Zdefn2=}
\end{eqnarray}
and hence from (\ref{Zdefn2}) the four new identities:
\begin{eqnarray}
\edth' \rho = - \rho\bar \tau+2\rho\tau'-\tau'\bar \rho,\nonumber\\
\thorn \tau' = - \rho\bar \tau+2\rho\tau'-\tau'\bar  \rho,
\label{Z_2}
\end{eqnarray}
together with their  $'$ counterparts.

\smallskip
The results so far obtained duplicate those of 
Carminati and Vu \cite{car1}; but we shall now continue the analysis even further.

\medskip

For consistency it is now necessary to substitute the value for $ Z_2$ into  (\ref{dZ2}); 
 making use of the Ricci equations, the post-Bianchi equations and the  identities (\ref{I1I2}), 
 yields the following linear equations for $Z_1$, $\bar Z_1$
\begin{eqnarray}
\bar \tau'  Z_1-\tau \bar Z_1=\tau\tau'\bar\tau' -\tau'\bar\tau'^2,
\label{1Z_1}\\
\bar\rho' Z_1 -\rho' \bar Z_1 = 2\rho'\bar\rho' (\rho-\bar \rho)+\tau'\bar\tau' (\bar\rho'-\rho')+\bar\rho'\Psi_2-\rho'\bar\Psi_2 \label{2Z_1},
\end{eqnarray}
together with the  $'$ counterpart  of (\ref{1Z_1}) (which is simply its complex conjugate), and the $'$ counterpart of (\ref{2Z_1}) given by
\begin{eqnarray}
\bar\rho Z_1 -\rho \bar Z_1 = 2\rho\bar\rho (\rho'-\bar \rho')+\tau\bar\tau (\bar\rho -\rho)+\bar\rho\Psi_2-\rho\bar\Psi_2 \label{3Z_1},
\end{eqnarray}
Note that (\ref{2Z_1}) and (\ref{3Z_1}) are not independent when $\rho'\ne 0 \ne \rho$, but in the cases $\rho'= 0$ or   $\rho =0$ they may provide different information.

In addition, it is also necessary to  apply each of the four operators to the identities (\ref{I1I2}) 
(in practice, only one of the operators needs to be checked because of symmetry);  the result is a duplication of the existing information.

\smallskip
\qquad  At this stage we have a complete and involutive set of tables for the six complex quantities in ${\cal S}^{+} = \{\rho, \tau, \rho', \tau', \Psi_2, Z_1\}$. However, this does not permit us yet to lower the Karlhede bound; in order to do that we  need to be able to substitute $Z_1$ in terms exclusively of the elements in ${\cal S}$.
\smallskip

Until now, we have been looking at the whole class of Petrov D vacuum spaces together: 
for $\Psi_2$ we have a {\it complete table of its derivatives}, i.e., expressions for  
all four of its GHP derivatives; but for each spin coefficient we have incomplete tables (from the point of view of elements of ${\cal S}$), 
since we have obtained explicit expressions built from ${\cal S}$, for only {\it three} of its derivatives. 
These expressions, contained in (\ref{ricci1})-(\ref{ricci4}), (\ref{Z_2}), involve only algebraic products of 
spin coefficients,
 and for 
the sake of clarity are gathered in the next expression
\begin{eqnarray}
\thorn\rho=\rho^2,\quad\edth\rho=\tau(\rho-\bar\rho),\quad\edth'\rho=-\rho\bar\tau+2\rho
\tau'-\tau'\bar\rho,\nonumber\\
\thorn\tau=\rho(\tau-\bar\tau'),
\quad\thorn'\tau=-\rho'\bar\tau'+2\rho'\tau-\tau\bar\rho',\quad\edth\tau=\tau^2 ; 
\label{n-form1}
\end{eqnarray}
From here we also get   the derivatives of $\rho'$, $\tau'$  by taking the prime of each equation.
There remains  one derivative for each spin coefficient which contains terms involving $Z_1$;  
 in the {\it generic case} (see {\bf Class IIIB}, below), we can obtain $Z_1$ from 
(\ref{1Z_1})-(\ref{3Z_1}), and  it is found to consist of only algebraic products of spin coefficients and $\Psi_2$,  but there are
 cases in which (\ref{1Z_1})-(\ref{3Z_1}) are not independent,  and hence we must proceed 
in a  different way. To
continue the analysis requires that we now separate into the different classes  which are presented in the Appendix.

\smallskip

\medskip
\noindent
{\em Class I}: \ $\rho \rho'\neq 0$; $\tau=0 = \tau'$.

Equation (\ref{Zdefn1})  implies $Z_1=0$ and hence adding (\ref{Zdefn1}) to  (\ref{n-form1}) it follows immediately that  {\it all the GHP 
derivatives of all non-zero elements of ${\cal S}$ are given by expressions constructed 
algebraically from  elements of ${\cal S}$:  a  complete involutive set of tables for the action of the GHP operators on all of the nonvanishing elements of ${\cal S}$.} 

Note that in this case we also have an extra relation involving the spin coefficients  which is obtained by setting $Z_1=0$ 
in 
(\ref{2Z_1}),
\begin{equation}
 \bar\Psi_2 \rho'+2
   (\bar\rho'-\rho')
   \bar\rho \rho'-\Psi_2 \bar\rho'=0,
\label{Psi2-extra}
\end{equation}
together with its  $'$ counterpart which is obtained from (\ref{3Z_1}). However, the latter is  not independent of (\ref{Psi2-extra}), since we assume $\rho'\ne 0 \ne \rho$ in this case. 

The integrability conditions arising from (\ref{n-form1}) give a set of conditions which depends 
functionally on (\ref{Psi2-extra}). Similarly if we differentiate (\ref{Psi2-extra}) and replace the 
derivatives of the spin coefficients and $\Psi_2$ by their expressions we get a condition which is functionally 
dependent on (\ref{Psi2-extra}) and thus gives nothing new.

\medskip
\noindent
{\em Class II}: \ $\tau\tau' \ne 0; \ \rho'=0 \ne \rho$.
 
The third line of (\ref{Zdefn1})  gives $Z_1=\Psi_2+\tau'\bar\tau'$ and  hence adding (\ref{Zdefn1}) to  (\ref{n-form1}) it follows immediately that all the GHP 
derivatives of all non-zero elements of ${\cal S}$ are given by expressions constructed 
algebraically from  elements of ${\cal S}$:  a  complete involutive set of tables for the action of the GHP operators on all of the nonvanishing elements of ${\cal S}$.

Also (\ref{1Z_1}) becomes
\begin{equation}
 \bar\Psi_2\tau+2
   (\tau-\bar\tau')\bar\tau\tau-\Psi_2\bar\tau'=0,
\label{cpsi1}
\end{equation}
but (\ref{2Z_1}) and (\ref{3Z_1}) are identically satisfied. 
The substitution of the values of $Z_1$ and $Z_2$ in (\ref{Zfode})-(\ref{dZ1}) yields the extra restriction 
(since $\rho \ne 0$)

\begin{equation}
 \rho(\Psi_2+\bar\Psi_2-2\tau\tau'+2\tau\bar\tau-2\bar\tau'\bar\tau)=0.
\label{cpsi2}
\end{equation}
The last pair of equations can be regarded as a linear system in the variables $\Psi_2$, $\bar\Psi_2$ from which it 
can be deduced that
\begin{equation}
(\tau + \bar\tau')\Psi_2= 2 \tau^2\tau'
\label{psibarpsi0}
\end{equation}
and  so 
\begin{equation}
\Psi_2= \frac{2 \tau^2\tau'}{\tau+\bar\tau'},
\label{psibarpsi}
\end{equation}
where, from (\ref{psibarpsi0}), $\tau'\neq 0$, $\tau\neq 0$ imply that $\tau+\bar\tau'\neq 0$. 

Next we insert (\ref{psibarpsi}) into (\ref{bianchi0})-(\ref{bianchi}) and their $\prime$ counterparts, 
and use the expressions
already found for the derivatives of the spin coefficients. The result is the restriction
\begin{equation}
 I_3 \equiv\rho\bar\tau+\bar\rho\tau'=0.
\label{restriction}
\end{equation}
Neither the integrability conditions of (\ref{n-form1})  nor the differentiation of (\ref{restriction}) give 
further conditions. Finally we note that the combination of (\ref{psibarpsi}) and (\ref{restriction}) yields the relation
\begin{equation}
\Psi_2\bar\rho^3+\bar\Psi_2\rho^3=0. 
\label{psirestriction}
\end{equation}

\smallskip

\medskip
\noindent
{\em Class II $'$}: \ $\tau\tau' \ne 0; \ \rho=0 \ne \rho'$.

This class does not need to be considered separately, since, by the interchange of the null vectors 
$l \leftrightarrow  n$, this class is transformed into the previous one. \smallskip

\medskip
\noindent
{\em Class IIIA}:    \ $\tau  \tau'\rho'\rho \ne 0$; $\bar\rho'\tau-\rho'\bar\tau'=0 = \bar\rho\tau'-\rho\bar\tau $. 

When we differentiate  the conditions $\bar\rho'\tau-\rho'\bar\tau'=0=\bar\rho\tau'-\rho\bar\tau $ we obtain
\begin{eqnarray} \rho=\bar\rho, \quad \rho'=\bar\rho',  \quad \tau=\bar\tau', 
 \quad \Psi_2=\bar\Psi_2,  \quad  Z_1=\bar Z_1,
\label{bar=}
\end{eqnarray}
We take as independent variables $\rho$, $\rho'$, $\tau$,  and $\Psi_2$.
Combining (\ref{bar=}) with (\ref{PsiZdefn2}) and (\ref{n-form1}) gives respectively 
\begin{eqnarray}
Z_2 =0 = Z_2',
\label{Z2=0}
\end{eqnarray}
and 
\begin{eqnarray}\thorn\tau = 0 = \thorn' \tau, \quad  \quad \edth \rho = 0 =  \edth \rho', \quad \edth' \rho' = 0 =  \edth' \rho . 
\label{ds=0}
\end{eqnarray}
The only remaining equations where we can get information about $Z_1$ are (\ref{Zfode})-(\ref{dZ1}), which simplify to
\begin{eqnarray}
\thorn Z_1 =  \rho ( Z_1  + \tau \bar\tau )  ,\label{Zfodes}\\
\edth Z_1 = 2\tau (Z_1 +\Psi_2  + \rho\bar\rho' ).  \label{dZ1ss} 
\end{eqnarray}
Putting
\begin{eqnarray}
 3A = (Z_1+\tau\bar\tau)/\Psi_2^{1/3}, \qquad \qquad \qquad 3B = \Psi_2^{1/3}+ 
(Z_1+\rho\bar\rho')/\Psi_2^{2/3},
\label{A,B}
\end{eqnarray}
we find directly from the above equations that
\begin{eqnarray}
\fl\thorn A = 0 = \thorn' A, \qquad\qquad\qquad\qquad\qquad\qquad  \thorn B = 0 = 
\thorn' B, \\
\fl\edth A = 2\tau B \Psi_2^{2/3}/3, \ \edth' A = 2\tau' B \Psi_2^{2/3}/3, \ \    \edth B = 3\tau  \Psi_2^{1/3}, \ \edth ' B = 3\tau'  \Psi_2^{1/3}.  
\label{dA,B}
\end{eqnarray}
From the last set of equations it follows that 
\begin{eqnarray}
 \edth (B^2-9A) = 0=  \edth' (B^2-9A).
\label{dA-B}
\end{eqnarray}
For zero-weighted quantities, such as $A, B$,  we have the relationship
$$
\nabla_a = n_a\thorn +l_a\thorn' -\bar m_a\edth- m_a \edth ',
$$
and so this implies, 
\begin{eqnarray}
 \nabla_{a} (B^2 -9A ) = 0 \quad    \Rightarrow \quad  (B^2 - 9A) = k, \ \hbox{ real constant}.
\label{nA-B}
\end{eqnarray}
Therefore 
\begin{eqnarray}(Z_1 + \Psi_2+\rho\bar\rho')^2 - 3(Z_1+\tau\bar\tau)\Psi_2     - k \Psi_2^{4/3} = 0,  \label{z1solution}
\end{eqnarray}
and so $Z_1$ can be obtained explicitly in terms of the  spin coefficients and $\Psi_2$ (and the constant $k$).

Therefore, by substitution for $Z_1$ from (\ref{z1solution}) into  (\ref{Zdefn1}),  and then  adding (\ref{Zdefn1}) to  (\ref{n-form1}), it follows immediately that  {\it all the GHP 
derivatives of all non-zero elements of ${\cal S}$ are given by expressions constructed 
 from  elements of ${\cal S}$:  a  complete involutive set of tables for the action of the GHP operators on all of the nonvanishing elements of ${\cal S}$.}

 The integrability conditions of this set of equations are identically fulfilled and thus they give no further restrictions.


\smallskip

\medskip
\noindent
{\em Classes IIIB,C}:    \ $\tau  \tau'\rho'\rho \ne 0$; $\bar\rho'\tau-\rho'\bar\tau'\ne 0 \ne
 \bar\rho\tau'-\rho\bar\tau $. 

In this case from (\ref{1Z_1})-(\ref{2Z_1}) it follows that
\begin{eqnarray}
(\rho' \bar \tau'-\bar \rho' \tau)Z_1 = 
\tau\Bigl(\bar\Psi_2\rho'+2 (\bar\rho'-\rho') 
\rho'\bar\rho-\Psi_2\bar\rho'
+(2\rho'\tau-\bar\rho'\tau
-\rho'\bar\tau')\bar\tau\Bigr).
\label{Z_1eqn2}
\end{eqnarray}
Given that $\bar\rho'\tau-\rho'\bar\tau' \ne 0$ this equation enables us to 
obtain $Z_1$,  and by inserting such a result in
(\ref{Zdefn1}), we obtain the expressions for the one remaining derivative of each spin coefficient. These expressions are 
\begin{eqnarray} 
\edthprime\tau &=& \frac{\tau\Bigl(2 \bar\rho
   \rho'^2-\bar\Psi_2\rho'+((\bar\tau'
-2\tau)\bar\tau-2\bar\rho'\bar\rho)\rho'+\Psi_2\bar\rho'
+\bar\rho'\tau \bar\tau\Bigr)}{\bar\rho'\tau-\rho'\bar\tau'},\nonumber\\
\thorn'\rho &=& \frac{\rho' \Bigl(\bar\Psi_2\tau+(\bar\rho'
   \bar\rho+2 (\tau-\bar\tau')\bar\tau) \tau-\Psi_2\bar\tau'+\rho'\bar\rho
 (\bar\tau'-2\tau)\Bigr)}{\rho'\bar\tau'
-\bar\rho'\tau}.
\label{TABLE}
\end{eqnarray}
together with their  $'$ counterparts. 

Hence,  adding (\ref{TABLE}) to  (\ref{n-form1}), it follows immediately that  {\it all the GHP 
derivatives of all non-zero elements of ${\cal S}$ are given by expressions constructed 
algebraically from  elements of ${\cal S}$:  a  complete involutive set of tables for the action of the GHP operators on all of the nonvanishing elements of ${\cal S}$.}

 The integrability conditions of this set of equations are identically fulfilled and thus they give no further restrictions.

\smallskip

\medskip
\noindent
{\em Class IV}:  \ $\tau\tau' \ne 0; \  \rho'=0=\rho$. 

The results from Class II are still valid and we only need to set 
$\rho=0$ in them; therefore {\it all the GHP derivatives of all non-zero elements of ${\cal S}$ are given by expressions constructed algebraically from  elements of ${\cal S}$: a complete involutive set of  tables for the action of the GHP operators on the non-vanishing  spin coefficients}. 

In this case, (\ref{cpsi2}) does not arise, and hence we only need to consider
(\ref{cpsi1}). Again, neither the differentiation of this condition nor the integrability conditions 
of (\ref{n-form1}) result in new relations.

\section{Invariant classification}
\label{classification}

The Karlhede algorithm \cite{kar} for an invariant classification of geometries  involves calculating --- in a particular frame ---
the  (zeroth derivative of)   Riemann tensor and then computing successively higher orders of its frame  covariant derivatives ('Cartan scalars'), 
together with possible changes of frame. The algorithm terminates at that particular order where (i) 
 no {\it new} functionally independent Cartan scalar is supplied by the frame derivatives of the Riemann tensor; \underbar{\it and} (ii) no {\it further fixing} 
of the frame is possible using these frame derivatives, i.e., the invariance group of the frame is unchanged.  This particular order is called the {\it Karlhede bound.}

Collins et al \cite{cdV1, cdV2} have used the GHP formalism to establish a relationship between the Cartan scalars and the GHP derivatives of the Weyl tensor, and hence provided an efficient  approach to the Karlhede classification of vacuum Petrov type D spacetimes. We shall now exploit their results directly,  and  strengthen their final conclusion. It is shown in  \cite{cdV1, cdV2}, and in their notation, that:

$\bullet$ the zeroth-order covariant derivative of the Weyl tensor is given  by $D^0 \Psi(\Psi_2)$; 

$\bullet$ the  first covariant derivatives of $\Psi_2$ are shown to be --- via the Bianchi equations ---  algebraic expressions involving only all elements of ${\cal S}$, and so are given  by $D^1 \Psi(\Psi_2,\rho,\rho',\tau,\tau')$; 
 for each of the different classes, the dimension of the invariance subgroup changes.

\smallskip
This means that we have to  continue to the next order of differentiation.  This is where we strengthen the earlier result because our analysis in the last section shows that derivatives of the spin coefficients do not in fact introduce any new quantities (since we have found explicit expressions for $Z_1$ exclusively in terms of the elements of ${\cal S}$,  in all cases.) Hence,

$\bullet$ the second covariant derivatives of $\Psi_2$ --- via Bianchi equations, Ricci equations, (\ref{n-form1}) and  (\ref{Zdefn1}) substituted  with the various explicit expressions for $Z_1$ ---  are   expressions involving  exclusively elements of ${\cal S}$  and are given by $D^2 \Psi(\Psi_2,\rho,\rho',\tau,\tau')$; also the dimension of the invariance subgroup does not change for any of the different classes.

(For completeness, we note that in \cite{cdV1} it was pointed out explicitly --- for the cases corresponding to our Classes I and IV --- that since all components of $D^2\Psi$ can be expressed algebraically in terms of first order quantities, the frame cannot be fixed any further at second order; in addition --- for the case corresponding to our Class III --- the frame was fixed completely at first order, and so no further fixing is possible at second order. 
For our  Class II --- which was overlooked in \cite{cdV1, cdV2} --- the frame can be fixed completely at first  order by choosing, e.g., $| \rho | = 1$  and $\tau = \bar\tau$; and so no further fixing is possible.) 
\smallskip

In summary,  all elements of ${\cal S}$ occur  at first derivative order, and no new quantities appear at second derivative order; moreover,  the invariance group remains unchanged from first to second derivative order: therefore the Karlhede algorithm terminates at second order,

\section{Conclusion and Discussion}

Thanks to the efficiency of the GHP formalism, we have rederived the identities (\ref{I1I2}), and derived  some new identities (\ref{1Z_1}), (\ref{2Z_1}), (\ref{3Z_1}),  (\ref{TABLE}) for the spin coefficients,   in a transparent manner; using these results it has been easy  to lower to {\it two} the upper bound for the invariant classification of vacuum Petrov D spaces.

In addition to showing in Section 2 that  all GHP derivatives 
of all non-zero elements in ${\cal S}$ are given by  expressions 
built exclusively from elements of ${\cal S}$,
we have obtained a  complete and involutive set of 
explicit tables for the action of all four GHP derivatives on the elements of ${\cal S}$ ---  in a manageable form, by a systematic and transparent analysis
 using the GHP formalism.  These identities  
(\ref{I1I2}), as well as {\it partial tables} (\ref{n-form1}) (for {\it three} GHP derivative operators acting on each spin coefficient) have already been obtained by Carminati and Vu \cite{car1} in the GHP formalism, but we believe that this is the first time that the spin coefficient identities (\ref{TABLE}) are presented and that all possible compatibility and integrability conditions have been explicitly checked for these spaces. These tables will be the basis for a systematic and efficient integration of the different type D vacuum classes in a future paper.   

In their NP Maple investigation \cite{czmc1}, Czapor and McLenaghan state that they have obtained the complete tables for 
all of the NP spin coefficients (including the 'badly-behaved' ones),  but that these tables, together with a lot of the intermediate calculations, were too long to print out explicitly in the NP formalism. Therefore, in their work \cite{czmc1}, there was perhaps the potential to deduce the result that the theoretical upper bound  was {\it two} in the invariant classification   (at least, in the generic case); however,  this possibility was obscured by the presence of the badly behaved spin coefficients which led to the very long and unmanageable expressions, and checking for possible tetrad changes would probably have been virtually impossible.  In comparison, the GHP calculations and results are manageable and transparent, and it is very instructive to compare how the calculations develop in the GHP procedure of this paper compared with the NP version in \cite{czmc1}.

The classification given in the Appendix improves on, and ties up a few loose ends regarding earlier classifications.
 
 \smallskip
 
The system {\em xAct} was used to cross-check all the computations derived by hand and to derive others which due to their complication were difficult to obtain without the aid of the computer. A {\em Mathematica} notebook in which all the equations of the GHP formalism are obtained from scratch is available in \cite{spinors}.   
 
\

\section{Acknowledgements}
AGP is supported by  the Spanish ``Ministerio de
Ciencia e Innovaci\'on'' under the postdoctoral grant
EX-2006-0092. He is also grateful to the Mittag-Leffler 
Institute, Djursholm, Sweden, for hospitality and financial 
support during part of the work on this paper.
JMM thanks Vetenskapsr\aa det (Swedish Research
Council) for supporting a visit to Link\"opings universitet.
He was also supported in part by the Spanish MEC Project
FIS2005-05736-C03-02. Both AGP and JMM thank the Department of Applied Mathematics of Link\"opings universitet, for their hospitality and support.

We thank Dr Lode Wylleman for a careful reading of this paper and for constructive comments.
 
\section*{Appendix:  Classification of vacuum Petrov type D spacetimes} 

We shall give a classification for vacuum Petrov type D spacetimes  which is well suited to our investigations; since this classification eventually 
turns out to be closely related to the Kinnersley classification we shall label it similarly to that classification.

In addition to the GHP zero-weighted quantities  $I_1$ and $I_2$ we shall use the GHP versions 
$$ I_3\equiv \bar\rho\tau'+\rho\bar\tau, \qquad  I_4\equiv \bar\rho'\tau+\rho'\bar\tau'$$
$$I_5\equiv \bar\rho\tau'-\rho\bar\tau, \qquad  I_6\equiv\bar\rho'\tau-\rho'\bar\tau'$$
of the NP quantities $I_3, I_4, I_5$ which have also been defined  in \cite{czmc2}, and in addition, for completeness, we have added $I_6$.

These are not invariants under tetrad transformations, but since they are of GHP weight, they are rather rescaled; so,  if for example $I_3=0$, then under arbitrary spin and boost tetrad transformations,  $I_3'=0$. By making direct use of the results $I_1=0=I_2$, the simple lemma follows.

\begin{lemma} If $\rho\ne 0$,  then:

 (i) $I_3=0\ \Rightarrow \ I_4=0$; 
 
 (ii)  $I_5=0\ \Rightarrow \ I_6=0$.
 
 \smallskip
 
  If $\rho'\ne 0$,  then:

 (i) $I_4=0\ \Rightarrow \ I_3=0$; 
 
 (ii)  $I_6=0\ \Rightarrow \ I_5=0$.

\end{lemma}

From the earlier work, we obtain  two other simple lemmas. 
\begin{lemma}
\begin{eqnarray}
(i) \ \rho=0 &\Rightarrow&  \tau \ne 0\ne\tau',\label{rholem1}\\
(ii) \ \rho'=0 &\Rightarrow&  \tau\ne  0\ne\tau'.\label{rholem2}
\end{eqnarray}
\label{rho-lemma}
\end{lemma}
{\em Proof:\hspace{2mm}} Assume that $\rho=0$; then  (\ref{ricci5}) becomes 
$$
0=- \tau \bar \tau -\Psi_2 + \edthprime\tau,
$$
which implies $\tau\neq 0$ in a type D spacetime. Combining this with the condition $I_1=0$ shown in (\ref{I1I2}) we conclude that $\tau'\neq 0$. If $\rho'=0$ the argument is similar. \qed

\smallskip

\begin{lemma}
$$(i) \ \rho\ne 0\ne \rho' , \tau=0\qquad\Rightarrow \qquad\tau'= 0$$
$$(ii) \ \rho\ne 0\ne \rho' , \tau'=0\qquad \Rightarrow\qquad  \tau= 0$$
\end{lemma}
{\em Proof:\hspace{2mm}} Apply (\ref{ricci3}) and its prime counterpart.
\qed

\medskip

From these lemmas it follows that any vacuum type D spacetime falls within one of the following mutually exclusive subclasses:

\smallskip
{\bf Class I.} $\rho\rho'\ne 0$; $\tau=0=\tau'$. 

\smallskip
{\bf Class II.}\footnote{This was the case overlooked in \cite{fla} and \cite{cdV1}, \cite{cdV2}.}   $\tau\tau'\ne 0$; $\rho\ne 0= \rho'$.\smallskip

[\ {\bf Class II$'$.} $\tau\tau'\ne 0$; $\rho'\ne 0= \rho$. \ ] 
\smallskip

{\bf Class III.} $\rho\rho'\tau\tau'\ne 0.$ \qquad 
\smallskip

{\bf Class IV.} $\tau\tau'\ne 0$; $\rho= 0= \rho'$. 

\medskip

{Class III}  can be further subdivided:

\smallskip
 {\bf Class IIIA.} $I_6= 0= I_5$; \ $I_4\ne 0\ne  I_3$.  

\smallskip
 {\bf Class IIIB.} $I_6\ne 0\ne I_5 $; \ $I_4\ne 0\ne  I_3$.

\smallskip
 {\bf Class IIIC.} $I_6\ne 0\ne I_5 $; \ $I_4 = 0 = I_3$.

\medskip

We have used these labels because of the substantial overlap with the Kinnersley cases: by examining the different explicit metrics obtained by Kinnersley, we deduce the following:
\begin{itemize}

 \item Class I coincides with Kinnersley Case I.

\item Class II coincides with  a subcase ($\rho'=0$) of Kinnersley Case IIE.  
\item Class IIIA coincides with  Kinnersley Case IIIA. 
\item Class IIIB  coincides with  Kinnersley Case IIIB. 
\item Class IIIC coincides with Kinnersley
IIA-D, F, together with  the other subcase ($\rho'\ne 0$) of Kinnersley Case IIE. 

\item Class IV coincides with Kinnersley Case IV.

\end{itemize}

A partial classification of the  vacuum Petrov D metrics with non-zero cosmological constant, using $I_1, I_2 \ldots I_5$, was given in  \cite{czmc2}, and the above results are  more complete.  Note that the table on p2166 of  \cite{czmc2} permits $\rho'=0\ne I_3$, which is ruled out above; and so   the only Kinnersley case with null orbits  (in the vacuum case with zero cosmological constant) is IIE. our attention has been drawn to a classification given in \cite{lode} which is very similar to the one in this paper, and it also notes these additional properties.

\end{document}